\documentclass[12pt]{article}

\usepackage{epsfig}

\def\bb{\mbox{\bf b}}
\def\bd{\mbox{\bf d}}

\def\bs{\mbox{\bf s}}
\def\bt{\mbox{\bf t}}
\def\bc{\mbox{\bf c}}
\def\bu{\mbox{\bf u}}

\def\bphi{\mbox{\boldmath $\phi$}}

\def\balpha{\mbox{\boldmath $\alpha$}}
\def\bbeta{\mbox{\boldmath $\beta$}}

\def\bPhi{\mbox{\boldmath $\Phi$}}

 \def\tfrac#1#2{{\textstyle{{#1}\over{#2}}}}

\def\half{\tfrac{1}{2}}
\def\third{\tfrac{1}{3}}

\def\twothirds{\tfrac{2}{3}}

\usepackage{amssymb}

\def\bbr{{\mathbb R}}

\begin{document}

\begin{titlepage}

\baselineskip 24pt

\begin{center}

{\Large {\bf Generation patterns, modified $\gamma-Z$ mixing, and hidden sector
with dark matter candidates as framed standard model results}} 

\vspace{.5cm}

\baselineskip 14pt

{\large Jos\'e BORDES \footnote{Work supported in part by Spanish
    MINECO under grant FPA2017-84543-P,
Severo Ochoa Excellence Program under grant SEV-2014-0398, and
Generalitat Valenciana under grant GVPROMETEOII 2014-049. }}\\
jose.m.bordes\,@\,uv.es \\
{\it Departament Fisica Teorica and IFIC, Centro Mixto CSIC, Universitat de 
Valencia, Calle Dr. Moliner 50, E-46100 Burjassot (Valencia), 
Spain}\\
\vspace{.2cm}
{\large CHAN Hong-Mo}\\
hong-mo.chan\,@\,stfc.ac.uk \\
{\it Rutherford Appleton Laboratory,\\
  Chilton, Didcot, Oxon, OX11 0QX, United Kingdom}\\
\vspace{.2cm}
{\large TSOU Sheung Tsun}\\
tsou\,@\,maths.ox.ac.uk\\
{\it Mathematical Institute, University of Oxford,\\
Radcliffe Observatory Quarter, Woodstock Road, \\
Oxford, OX2 6GG, United Kingdom}

\end{center}

\vspace{.3cm}

\begin{abstract}

A descriptive summary is given of the results to-date from the framed standard 
model (FSM) which:
\begin{itemize}
\item assigns geometric meaning to the Higgs field and to fermion generations,
hence offering an explanation for the observed mass and mixing patterns of 
quarks and leptons, reproducing near-quantitatively 17 of SM parameters with
only 7.
\item predicts a new vector boson $G$ which mixes with $\gamma$ and $Z$, leading
to deviations from the SM mixing scheme.  For $m_G > 1$ TeV, these deviations
are within present experimental errors but should soon be detectable at LHC 
when experimental accuracy is further improved.
\item suggests the existence of a hidden sector of particles as yet unknown to 
experiment which interact but little with the known particles.  The lowest 
members of the hidden sector of mass around 17 MeV, being electrically neutral 
and stable, may figure as dark matter constituents.
\end{itemize}  
The idea is to retrace the steps leading to the above results unencumbered by 
details already worked out and reported elsewhere.  This has helped to clarify 
the logic, tighten some arguments and dispense with one major assumption 
previously thought necessary, thus strenthening earlier results in
opening up possibly a new and
exciting vista for further exploration.

\end{abstract}

\end{titlepage}

\clearpage

The FSM was initially conceived to address the generation puzzle but,  
for its own consistency, has led to consequences way beyond its original 
remits.  To trace how this comes about is the aim of the present article.

\section{Generation puzzle}

By the generation puzzle we mean the following empirical facts \cite{pdg}:
\begin{itemize}
\item  Quarks and leptons occur in 3 generations.
\item  Generations have hierarchical masses, e.g. $m_t \gg m_c \gg m_u$.
\item  Up and down flavoured states are not aligned giving e.g. for quarks:
\begin{equation}
V_{\rm CKM} = \left( \begin{array}{ccc} 
                              \bu \cdot \bd & \bu \cdot \bs & \bu \cdot \bb \\
                              \bc \cdot \bd & \bc \cdot \bs & \bc \cdot \bb \\
                              \bt \cdot \bd & \bt \cdot \bs & \bt \cdot \bb
         \end{array} \right)
         \neq I. 
\label{CKM}
\end{equation}
\end{itemize}
For these one would wish to have an explanation \cite{weinberginter}.
The standard model (SM)
takes these facts for granted, which account for some two-thirds of its many 
empirical parameters.

\section{Rotating rank-one mass matrix}

Towards understanding the generation puzzle, one first step made some years
ago was R2M2 (Rotating Rank-one Mass Matrix) \cite{phenodsm,r2m2,Bjwebsite} 
which shows that if: 
\begin{itemize}
\item the quark and lepton mass matrices are of the common form:
\begin{equation}
m = m_T \balpha \balpha^\dagger
\label{mfact}
\end{equation}
where $\balpha$, a vector in 3D generation space, is the same for all quarks
and leptons and only the numerical coefficients $m_T$ differ, and
\item $\balpha$ rotates as the scale changes,
\end{itemize}
then the above features can be qualitatively reproduced.

We know that coupling constants and masses can change with scale as a result 
of renormalization.  Similarly, renormalization can cause the mass matrix which 
has an orientation (in generation space) to rotate with changing scales.  
Indeed, even in the usual formulation of the SM, the mass matrix rotates as 
a result of mixing \cite{smrotate}.  The new point  in R2M2 then is 
that it is the rotation which gives rise to the mixing rather than the other 
way round.  

That R2M2 will give rise to mixing is easy to see. A mass matrix rotating with 
scale means that not only its eigenvalues but also its eigenvectors will be 
scale-dependent.  Now the masses and state vectors of particles are supposed 
each to be measured at their respective mass scales.  Since, for example, the 
up-type quarks $t, c, u$ have different masses from the down-type quarks 
$b, s, d$, it follows that the state vectors of the two types, say respectively 
$\bt, \bc, \bu$ and $\bb, \bs, \bd$, which are eigenvectors of the matrix at 
their own different mass scales will also be different.  Hence, the CKM matrix
\cite{CKM} (\ref{CKM}) will not be the identity matrix which is what is meant 
by mixing.

That R2M2 will lead to a hierarchical mass spectrum is also quite easy to see.
The matrix (\ref{mfact}) has only one nonzero eigenvalue, namely $m_T$ with
eigenvector $\balpha$, which, for example, at scale $\mu = m_t$ gives $\bt 
= \balpha(m_t)$.  The two other eigenvectors have zero eigenvalues at this 
$\mu$, but these are not the masses of the two lower generations $c$ and $u$ 
for these mass values have to be measured at their own mass scales.  What 
happens at these lower scales is that because of the rotation of $\balpha$, 
there is some ``leakage'' of $m_t$ to the lower generations $c$ and $u$ to give 
each a mass, hence mass hierarchy.

Closer examination of R2M2, in fact, reveals many other detailed features which 
are also seen in experiment
\cite{phenodsm,features,strongcp,atof2cps,cornerel}.

\section{Framing SM for FSM}

The FSM is an extension of the SM constructed to give R2M2 as a result.  We 
recall that the standard model is a gauge theory based on the gauge symmetry 
$G = U(1) \times SU(2) \times SU(3)$, with the gauge (vector boson) fields, and 
the matter (fermion) fields as dynamical variables, to which is added
a scalar Higgs field to break the $SU(2)$ symmetry, as demanded
by experiment.  The FSM is an extension of this set-up.  However, in contrast to
usual extensions met in the literature, such as GUT or SUSY, it does not seek 
an enlargement of the (local) gauge symmetry.  It keeps the same $G$ as the SM,
but instead adds to the SM as new dynamical variables the frame vectors in the 
internal symmetry space, as follows.

We recall that a gauge theory is by definition invariant under local gauge 
transformations.  These transformations can be and are usually represented as 
matrices relating the local (spacetime point $x$-dependent) frame to a global
(spacetime point $x$-independent) reference frame.  The columns of such matrices are
often referred to as frame vectors.  The suggestion of FSM  is to add the 
elements of these transformation matrices or of the frame vectors as dynamical 
variables to the usual SM set.

Promoting frame vectors to dynamical variables is not a new idea.  General 
relativity is usually formulated in terms of the metric tensor $g_{\mu\nu}$ as 
dynamical variables, but can alternatively be formulated (Einstein--Cartan theory) 
\cite{vierbein} in terms of vierbeins, $e_\mu^a$, which are frame vectors in 
the language of the preceding paragraph, with $\mu$ referring to the local 
(co-ordinate) frame and $a$ referring to the global reference frame, and the 
metric tensor $g_{\mu\nu}$ appears then as:
\begin{equation}
g_{\mu\nu} = \sum_a e^a_\mu e^a_\nu.
\label{gmunu}
\end{equation}
So the suggestion of making frame vectors into dynamical variables in FSM 
makes, in a sense, the particle theory closer in spirit to the theory of 
gravity \cite{efgt} and might make it easier in  future for the eventual 
unification of the two theories.

Frame vectors as dynamical variables, or framons as we shall call them, have
a special property that the gauge bosons and matter fermions do not have.
Being transformation matrices between the local frame and a global reference 
frame, they depend naturally on both the local and global frames, and transform
when either the local or the global reference frame is changed.  This is like
the vierbeins in gravity, which is seen above to carry two indices, one local 
and one global, but unlike the gauge bosons and matter fermions which transform 
only under local gauge transformations.  Since physics should be invariant 
under changes in both the local and global reference frame, it means that for 
a theory with the SM gauge symmetry $G = U(1) \times SU(2) \times SU(3)$, the 
action when framons are included as dynamical variables should be invariant 
under not only (local) $G$ but also a global $\tilde{G} = \tilde{U}(1) \times 
\widetilde{SU}(2) \times \widetilde{SU}(3)$.  This additional requirement does 
not affect those terms originally present in the SM involving only the gauge 
bosons and matter fermions fields, since these fields are themselves invariant 
under $\tilde{G}$, but will put restrictive constraints on the new terms 
involving the new framon fields, and help to specify the theory.

The SM gauge symmetry $G = U(1) \times SU(2) \times SU(3)$ being a product of
three simple symmetries (modulo some discrete identifications which do not 
conern us here), there are several different ways to represent the gauge 
transformations as matrices, depending on whether for the product of each pair
we take the sum or the product representation.  To fit the SM framework, the 
FSM \cite{efgt,dfsm,tfsm} opted for the choice:
\begin{equation} 
{\bf 1} \times ({\bf 2} + {\bf 3})
\label{framonrepl}
\end{equation}
for the local $G$, where ${\bf 1}$ is for $U(1)$, ${\bf 2}$ means the doublet
for $SU(2)$ and ${\bf 3}$ the triplet for $SU(3)$, but
\begin{equation}
{\bf \tilde{1}} \times {\bf \tilde{2}} \times {\bf \tilde{3}}
\label{framonrepg}
\end{equation}
for the global $\tilde{G}$.  This choice is the closest, in fact the only one
close, to the SM in structure and happens also to require the smallest number 
of framon fields to be introduced \cite{dfsm}.  
 
Specifically, this means that the framon matrix breaks up into two parts:
\begin{itemize}
\item {\bf [FF]} the ``flavour framon'': $\balpha \Phi$,
\item {\bf [CF]} the ``colour framon'': $\bbeta \bPhi$.
\end{itemize}
The factors $\balpha$ and $\bbeta$ are global (space-time $x$-independent) 
quantities, where $\balpha$ transforms as a triplet under $\widetilde{SU}(3)$, 
and $\bbeta$ transforms as a doublet under $\widetilde{SU}(2)$.  Next, $\Phi$ 
is a scalar field (space-time scalar $x$-dependent quantity), as well as a 
$2 \times 2$ matrix whose rows transform as (local flavour) $SU(2)$ doublets 
but whose columns as (global flavour) $\widetilde{SU}(2)$ anti-doublets.  
Similarly, $\bPhi$ is a scalar field as well as a $3 \times 3$ matrix whose 
rows transform as (local colour) $SU(3)$ triplets but whose columns as (global 
colour) $\widetilde{SU}(3)$ anti-triplets.

The two columns of $\Phi$ are flavour doublet scalar fields either of which 
can play the role of the standard Higgs field in the electroweak theory.  The
SM, however, requires only one such.  To conform with this requirement,
the FSM imposes the following orthonormal condition:
\begin{equation}
\bphi_r^{\tilde{2}} \, = \, - \,  \epsilon_{rs} (\bphi_s^{\tilde{1}})^*
\label{minemb}
\end{equation}
on $\Phi$ so that one column can be eliminated leaving only the other to be
identified with the standard Higgs field.  This reduces further the number of
scalar fields introduced by framing, in the same spirit as the ``minimal'' 
choice (\ref{framonrepl}) the framon representation.  Now the possibility to 
impose such a condition on the framon field is unique to the group $SU(2)$, 
not adaptable to colour $SU(3)$ nor to $SU(N)$ for any larger
$N$, having to do with 
the possibility of embedding $SU(2)$ in ${\bbr}^4$ \cite{dfsm}, and 
(\ref{minemb}) will for this reason be referred to as the minimal embedding 
condition. 

\section{First FSM results}

Once this form for the framon is written down, several advantages of the FSM 
scheme become immediately apparent.
\begin{itemize}
\item The standard Higgs field in electroweak theory being now identified as a 
column of the framon $\Phi$ is thus given a geometric significance (namely as a
frame vector) which is missing in the usual formulation of the SM.
\item There has appeared a global 3-fold symmetry $\widetilde{SU}(3)$ which
can be taken as fermion generations, and if so gives to the latter also a 
geometric significance (as the ``dual'' to colour) which is missing also
in the usual formulation.
\item The mass matrices of quarks and leptons appear automatically in the form
(\ref{mfact}), where $\balpha$ can be taken here without any loss of generality 
as a real unit vector \cite{dfsm}.  Coming from the flavour framon {\bf [FF]}
playing the role of the Higgs scalar, $\balpha$ is naturally independent of the 
type of quark or lepton to which the mass matrix refers, as is wanted in R2M2.
\item The colour framon {\bf [CF]} carries both local colour and global colour
(generation) indices and, when appearing in loop diagrams, automatically 
generates rotation of $\balpha$ with changing scales, as is wanted in R2M2 
to give hierachical mass and mixing patterns.
\end{itemize}
The first two items are bonuses while the last two are the stated aims of the
FSM.  But will the scheme really work as hoped?  

The requirement of the doubled invariance under $G \times \tilde{G}$ already 
mentioned restricts the framon action sufficiently for simple loop diagrams 
to be calculated, and so the question posed can immediately be put to the 
test.  This is done in \cite{tfsm} where a renormalization group equation 
(RGE) for the scale dependence of $\balpha$ to one-framon-loop is derived.  
The equation itself depends on some parameters and when applied, depends on 
some integration constants, making it 7 real adjustable parameters in all.  
This is then required to fit experiment and Table \ref{tfsmfit} is obtained 
as the result.  One sees there that the masses and mixing elements measured 
in experiment have all been fitted quite well, with most fitted within 1.5 
$\sigma$ and none too wild, despite their intricate variations in size 
over a wide range.\footnote{This means effectively that the FSM has, to this 
accuracy, replaced by 7 adjustable parameters 17 of the SM's empirical 
parameters, although not all of these latter have been measured.}  

\begin{table}
\centering
\begin{tabular}{|l|l|l|l|l|}
\hline
& Expt (June 2014) & FSM Calc & Agree to & Control Calc\\
\hline
&&&& \\
{\sl INPUT} &&&&\\
$m_c$ & $1.275 \pm 0.025$ GeV & $1.275$ GeV & $< 1 \sigma$&$1.2755$ GeV\\
$m_\mu$ & $0.10566$ GeV & $0.1054$ GeV & $0.2 \%$ & $0.1056$ GeV\\
$m_e$ & $0.511$ MeV &$0.513$ MeV & $0.4 \%$ &$0.518$ MeV\\
$|V_{us}|$ & $0.22534 \pm  0.00065$ & $0.22493$ & $< 1 \sigma$ &$0.22468$\\
$|V_{ub}|$ & $0.00351^{+0.00015}_{-0.00014}$& $0.00346$ & $< 1 \sigma$&$0.00346$ \\
$\sin^2 2\theta_{13}$ & $0.095 \pm 0.010$ & $0.101$ &$< 1 \sigma$ &$0.102$\\
\hline
&&&& \\
{\sl OUTPUT} &&&&\\
$m_s$ & $0.095 \pm 0.005$ GeV & $0.169$ GeV & QCD &$0.170$ GeV \\
& (at 2 GeV) &(at $m_s$) &running& \\
$m_u/m_d$ & $0.38$---$0.58$ & $0.56$ &  $< 1 \sigma$&$0.56$ \\
$|V_{ud}|$ &$0.97427 \pm 0.00015$ & $0.97437$ & $< 1 \sigma$&$0.97443$ \\
$|V_{cs}|$ &$0.97344\pm0.00016$ & $0.97350$ & $< 1 \sigma$&$0.97356$ \\
$|V_{tb}|$ &$0.999146^{+0.000021}_{-0.000046}$ & $0.99907$ &$1.65
\sigma$&$0.999075$ \\
$|V_{cd}|$ &$0.22520 \pm 0.00065$ & $0.22462$ & $< 1 \sigma$ &$0.22437$\\
$|V_{cb}|$ & $0.0412^{+0.0011}_{-0.0005}$ & $0.0429$ & $1.55 \sigma$&
$0.0429$ \\
$|V_{ts}|$ & $0.0404^{+0.0011}_{-0.0004}$ & $0.0413$ &$< 1 \sigma$& 
$0.0412$\\  
$|V_{td}|$ & $0.00867^{+0.00029}_{-0.00031}$ & $0.01223$ & 41 \% & $0.01221$\\
$|J|$ & $\left(2.96^{+0,20}_{-0.16} \right) \times 10^{-5}$ & $2.35
\times 10^{-5}$ & 20 \% &$2.34\times 10^{-5}$ \\
$\sin^2 2\theta_{12}$ & $0.857 \pm 0.024$ & $0.841$ &  $< 1 \sigma$& $0.840$\\ 
$\sin^2 2\theta_{23}$ & $>0.95$ & $0.89$ & $> 6 \%$ &$0.89$\\
\hline 
\end{tabular}
\caption{Calculated fermion masses and mixing parameters compared with
experiment, reproduced from \cite{tfsm}} 
\label{tfsmfit}
\end{table}

As already noted, R2M2 being already incorporated, FSM can be expected to give 
the qualitative features correctly.  But this does not by any means guarantee 
that the values of the mass and mixing parameters can be correctly reproduced.
For these, the details of the rotation trajectory of $\balpha$ matter, namely
the shape of the curve it traces on the unit sphere as well as the variable
speed with respect to change of scale at which it moves along this curve, and 
it looks nontrivial that the FSM seems to have got it right. 

\begin{figure}
\centering
\includegraphics[height=17cm]{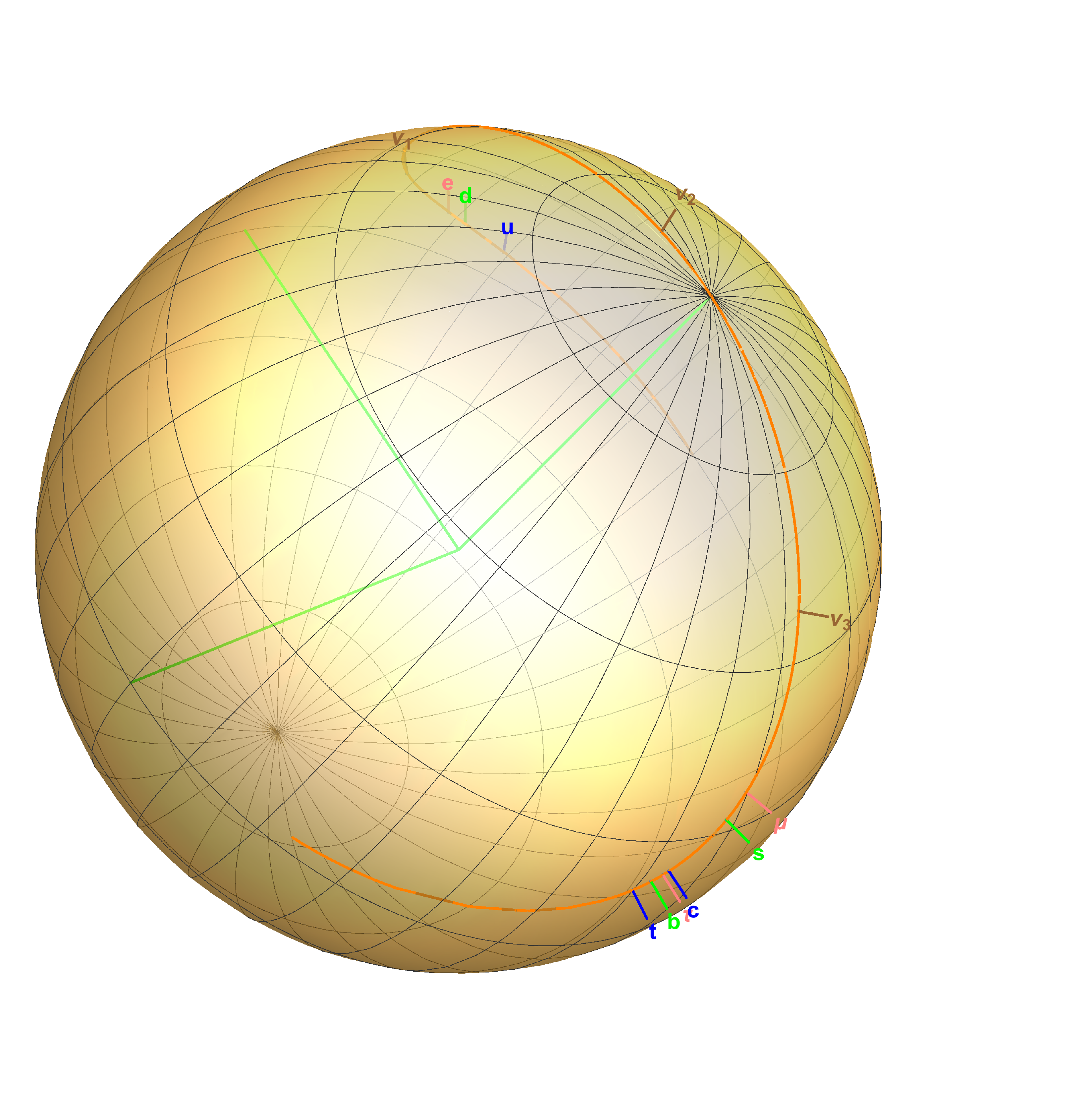}
\caption{The rotation trajectory of $\balpha$ on the unit sphere as determined
by the fit in \cite{tfsm}.  Note the change in sign of normal curvature around
the pole.}
\label{florosphere}
\end{figure}

In relation to Table \ref{tfsmfit} two points are particularly noteworthy:
\begin{itemize}
\item {\bf [a]} 
The QCD action is well known to admit CP-violation via a so-called theta-angle 
term \cite{weinbergbook} of topological origin which, if admitted with $\theta$
naturally of order unity, would lead to CP-violations in strong interactions 
many orders above what is seen in, for example, the neutron dipole moment.  It 
is also well known that this so-called strong CP problem can be solved if the 
quark mass matrix has zero eigenvalues, but this seems to contradict the 
empirical observation that all known quarks have nonzero mass.  Now it happens 
that in R2M2, and hence also the FSM, the quark mass matrix (\ref{mfact}) does 
have zero eigenvalues but yet all quarks have finite masses by virtue of the 
``leakage mechanism'' already mentioned, so that the strong CP-problem can be
solved as above by transforming away the theta-angle term.  However, the 
effect of eliminating the theta-angle term is transmitted by the rotation of
$\balpha$ to the CKM matrix to give it a Kobayashi-Moskawa CP-violating phase. 
Moreover, a $\theta$ of order unity is shown to give a KM phase or Jarlskog
invariant \cite{jarlskog}
of the right size \cite{strongcp,atof2cps}.  In other words, the FSM offers
a simultaneous solution to both the strong CP-problem and to the question why a 
KM phase of a certain size should appear in the CKM matrix for
quarks.  \footnote{It is known that in  the weak lagrangian a similar
topological term can be rotated away without any physical
consequences.  Hence if CP violation in the leptonic sector were also
due to a topological term as in QCD, then it would seem to indicate
that there is no CP violating Dirac phase in the PMNS matrix for leptons \cite{PMNS}.
However, there may be
other sources of CP violation, and in any case we know that there is
the possibility of CP violation due to Majorana phases.} 
\item {\bf [b]}
It is a crucial empirical fact that $m_u < m_d$, which is what makes the proton
lighter than the neutron and therefore stable, or otherwise we ourselves would
not be here.  But this looks anomalous, given that the up-type quarks of the 
heavier generations are heavier than their down-type counterparts, namely 
$m_t \gg m_b$ and $m_c > m_s$.  The FSM fit in Table \ref{tfsmfit}, however, 
gives the right answer $m_u < m_d$, indeed even to the ratio $m_u/m_d$.  This 
comes about as follows.  In Figure \ref{florosphere} is shown the trajectory of 
$\balpha$ on the unit sphere as obtained in the FSM fit of \cite{tfsm}.  There 
is a change in the (normal) curvature between $c, s$ and $u, d$, which is what 
gives in \cite{tfsm} this ``anomaly''.  It is a special intrinsic property of 
the RGE for $\balpha$ derived from the FSM which cannot be envisaged from 
[R2M2] alone, and is to play another significant role later.
\end{itemize}

\section{Questions of consistency}

Taken in all then, the FSM seems to have done the job it was intended to do
in reproducing the empirical mass and mixing patterns of quarks and leptons.
Even if taken just as a parametrization of the data, it is competitive with
any seen in the literature.  But this is not enough.  In constructing the FSM, 
new assumptions have been made, and new assumptions imply new physics.  One is 
obliged therefore to ascertain whether the new physics predicted by the FSM is 
consistent with present experiments, and if so, whether it can be checked 
further by experiments in future.

Now the new ingredients introduced by FSM are the framons, of which the flavour 
framon {\bf [FF]} has already been identified with the standard Higgs scalar. 
So what is newly introduced by the FSM over the SM are just the colour framons 
{\bf [CF]}.  But these  represent 9 new complex degrees of freedom.  Then
why have we not been made aware of them?  This is not immediately answerable 
because the colour framons are coloured and colour is confined, so that colour 
framons cannot propagate freely as particles in space.  They can however 
combine with one another and with other coloured objects via colour confinement 
into colour neutral bound states, and these can appear then as particles.  A 
framon can combine with an antiframon in $s$-wave to form a colourless scalar 
bound states which we shall call generically $H$, or in $p$-wave to form vector 
bound states which we call generically $G$, which via the covariant derivative 
will bring in the colour gluon.  Or a framon can combine with a coloured 
fermion to form fermionic bound states which we call generically $F$.   In this 
case, an immediate question is:
\begin{itemize}
\item {\bf [Q1]} Why have we not seen these particles $H, G, F$?
\end{itemize}
This is a question that FSM has to answer to remain viable, but it can only do
so when enough has been learnt about the properties of $H, G$ and $F$. 

To this end, let us ask ourselves another question which needs to be asked for 
internal consistency in any case, namely 
\begin{itemize}
\item {\bf [Q2]} In the flavour theory, the introduction of the Higgs scalar 
(the flavour framon here) with non-zero vacuum expectation value
breaks the $SU(2)$ gauge symmetry 
and gives masses to the quarks, the leptons, the vector bosons $W, Z$ and the 
scalar Higgs boson $h$.  By analogy then, why does the introduction of the 
colour framon, also with nonzero vacuum expectation value
to break the generation symmetry, not 
break the colour gauge symmetry and give massive fermions, vector bosons and 
Higgs scalar bosons in analogy to the above?
\end{itemize}
At first sight, this seems a totally unrelated question but it will soon be seen
 to be just another way of posing {\bf [Q1]}, and that the recognition of this
equivalence will lead us a long way towards understanding the properties of the
$H, G, F$ that we seek.  

That {\bf [Q2]} is indeed equivalent to {\bf [Q1]} can be seen by a deep and 
subtle fact which has taken the perspicacity of 't Hooft \cite{tHooft} first
to point out.  In this illuminating paper, 't Hooft, among other things, made 
the observation that the standard (Salam-Weinberg) electroweak theory, which is
usually said to have its flavour $SU(2)$ symmery spontaneously broken, has a
``mathematically equivalent'' interpretation as a theory in which the $SU(2)$ 
theory is confining and exact, what is broken being only a global (spacetime 
$x$-independent) symmetry, say $\widetilde{SU}(2)$, associated with it which 
the electroweak theory is known to possess (called ``accidental'' by some).  
In this alternative interpretation, which we shall call the confinement picture,
the Higgs boson $h$ appears as a flavour neutral bound state of the original
flavoured Higgs scalar field $\phi$ with its conjugate $\phi^\dagger$ in the
$s$-wave, the massive vector bosons $W, Z$ appear as the same but now in the 
$p$-wave, while the quarks and leptons appear as flavour neutral bound states 
of $\phi$ with the fundamental flavour doublet fermion fields.  This means,
first, that the presence of scalar fields with nonzero vacuum
expectation value does not by itself
preclude the theory being confining, 
answering thus the first half of {\bf [Q2]}.  Secondly, we see that as above 
interpreted in the confinement picture, the $h$, $W, Z$ and $q, \ell$ in the 
flavour theory would be the exact analogues of respectively the $H$, $G$, and 
$F$ in the colour theory, only with flavour and colour interchanged.  Or, in 
other words, {\bf [Q2]} is just {\bf [Q1]} rephrased, as claimed.

\section{Flavour-colour parallel and the dichotomy of matter}

Now, for the FSM, 't Hooft's confinement picture for the electroweak theory
is a veritable eye-opener in that it reveals a very close parallel between the
two nonabelian component theories of the FSM, a parallel that was at first
entirely unsuspected.  This is schematised in Figure \ref{fvsc}, where we see
that both flavour and colour are now confined and both of them are framed by 
scalar framon fields.  Besides, both the flavour and colour framons have 
nonzero vacuum expectation values, which lead to the breaking of both the 
global symmetries, first of the $\widetilde{SU}(2)$ symmetry giving the two 
up-down flavours, and second, of the $\widetilde{SU}(3)$ symmery giving the 
three generations.  The cited vacuum expectation value
of the flavour framon (that is the Higgs 
scalar) is well known, while the cited vacuum expectation value
of the colour framon will appear 
later on.  Then the flavour framon combines with its own conjugate and with 
flavoured fermions via flavour confinement to form the particles $h, W, Z, q, 
\ell$ on the left, while the colour framons combine parallelly via colour 
confinement to form the particles $H$, $G$, and $F$ on the right.

\begin{figure}
  \centering
\setlength{\unitlength}{1.16mm}
\begin{picture}(140,140)
\put(40,0){\line(0,1){140}}
\put(90,0){\line(0,1){140}}  
\put(0,0){\line(1,0){140}}
\put(0,130){\line(1,0){140}}
\put(0,120){\line(1,0){140}}
\put(42,133){\bf Flavour Theory}
\put(92,133){\bf Colour Theory}
\put(0,123){Gauge Symmetry (local)}
\put(42,123){$SU(2)$}
\put(92,123){$SU(3)$}
\put(0,113){Confinement?}
\put(42,113){Confined, exact}
\put(92,113){Confined, exact}
\put(45,109){('t Hooft's picture)}
\put(95,109){(general consensus)}
\put(0,106){\line(1,0){140}}
\put(0,99){Framon scalar}
\put(0,92){\line(1,0){140}}
\put(42,99){$\Phi$\quad (flavour framons}
\put(51,95){$\Rightarrow$\ standard Higgs)}
\put(92,99){$\bPhi$\quad (colour framons}
\put(101,95){{\sl new} for FSM)}
\put(0,85){Symmetry doubled}
\thicklines
\put(13,96){\vector(0,-1){8}}
\put(47,85){$SU(2)\ \times\ \widetilde{SU}(2)$}
\put(42,76){{\sl local}}
\put(70,76){{\sl global}}
\put(97,85){$SU(3)\ \times\ \widetilde{SU}(3)$}
\put(92,76){{\sl local}}
\put(120,76){{\sl global}}
\thinlines
\put(50,84){\vector(-1,-1){5}}
\put(100,84){\vector(-1,-1){5}}
\put(68,84){\vector(1,-1){5}}
\put(118,84){\vector(1,-1){5}}
\put(0,73){\line(1,0){140}}
\put(0,66){Framon vev $\not= 0$}
\put(42,66){$\zeta_W$ (246 GeV)}
\put(92,66){$\zeta_S$ ($\sim$ TeV)}
\put(0,60){\line(1,0){140}}
\put(0,53){Global symmetry broken}
\thicklines
\put(13,64){\vector(0,-1){8}}
\put(47,52){$\widetilde{SU}(2)$ broken}
\put(42,48){$\Rightarrow$ up-down flavour}
\put(97,52){$\widetilde{SU}(3)$ broken}
\put(92,48){$\Rightarrow$ 3 generations}
\thinlines
\put(0,43){\line(1,0){140}}
\put(0,36){Framon bound states}
\put(41,36){by $SU(2)$ confinment ($\!$'t Hooft)$\!$}
\put(91,36){by colour $SU(3)$ confinement}
\put(42,27){$\to h, W,Z, \ q\ ,\ell$}
\put(61,28){\circle{4}}
\put(92,27){$\to H,G,F$}
\put(109,29){\vector(2,1){5}}
\put(109,27){\vector(2,-1){5}}
\put(117,30){$Q$ \ co-quarks}
\put(118.3,31){\circle{5}}
\put(117,23){$L$ \ co-leptons}
\put(0,18){\line(1,0){140}}
\put(0,13){Higher level bound states}
\put(42,13){$Q\bar{Q}, QQ$ co-hadrons}
\put(62,25.5){\vector(3,-1){33}}
\put(47,8){by flavour confinement}
\put(96,13){$q \bar{q}, qqq$ hadrons}
\put(101,8){by colour confinement}
\put(116,28){\vector(-3,-1){42}}

\end{picture}
\caption{Comparing the flavour and colour theories in FSM}
\label{fvsc}
\end{figure}

At first sight, this close parallel between the two sides may seem worrisome, 
since we are used to the conception of the flavour theory as describing the 
weak interactions and the colour theory as describing the strong.  But this
conception is a little loose and needs to be scrutinized anew in the present
context. When we said above that flavour interactions are weak while colour 
interactions are strong, what we really meant was that the particles bound by 
flavour confinement that we know, namely what we shall call for simplicity 
the ``weak particles'' $h$, $W, Z$, and $q, \ell$ interact weakly while 
those particles bound by colour confinement that we know, namely the hadrons, 
interact strongly.  But this is not comparing like with like in terms of the 
parallel set out in Figure \ref{fvsc}.  The analogues of $h$, $W, Z$ 
and $q, \ell$ on the left, we agreed, are the $H$, $G$, and $F$ on the right, 
{\em not} the hadrons.  Hadrons are constructs of a very different sort.  They 
are bound states, indeed also by colour confinement as are the $H, G, F$, but 
not of a colour framon with its conjugate or with something else, but of quarks 
and antiquarks which, though coloured, are themselves already flavour neutral
composites of a flavour framon bound to other flavoured constituent via flavour 
$SU(2)$ confinement.  Hadrons are thus, in a sense, higher-level constructs, 
{\em not} the parallel of the weak particles $h$, $W, Z$, and $q, \ell$.  The correct 
parallel of hadrons instead would be flavour neutral bound states by flavour 
confinement of some $F$s carrying flavour, if such exist, which we may call 
co-quarks $Q$ and anti-co-quarks $\bar{Q}$, resulting in say $Q \bar{Q}$ 
or $QQ$ which we may call co-hadrons (co-mesons or co-baryons).  The parallel 
between the left- and right-hand sides of Figure \ref{fvsc} is still maintained 
so long as the $H$, $G$, and $F$ are point-like and weakly interacting like the
weak particles, as will be shown later to be the case, while co-hadrons can
be bulky and strongly interacting, like hadrons.  From the perspective of Figure
\ref{fvsc}, what gives us the wrong impression that flavour interactions are 
weak and colour strong is just that of the bound states confined by flavour,
we know in experiment only the weak particles $h$, $W, Z$, $q, \ell$, and of
the bound states confined by colour, we know in experiment only the hadrons.
In other words, what remains a mystery is still just {\bf [Q1]}, and so long
as that is understood, as we hope to do later, the parallel exhibited in Figure
\ref{fvsc} holds good. 
 
Obviously, such a close parallel between the flavour and colour sectors in the 
FSM will have far-reaching consequences.  First, it leads to a conception which 
we might call:  
\begin{itemize}
\item {\bf [DoM]} {\it The flavour-colour dichotomy of matter}, where the
material world is partitioned into two sectors related to each other by 
having the roles of flavour and colour interchanged.  One (the standard sector)
already known to us is composed of the weak particles $h$, $W, Z$ and $q, 
\ell$ as building blocks, while the other (the ``hidden'' sector) is composed 
of their colour analogues $H$, $G$, and $F$ as building blocks.  From these
building blocks on either side, higher-level constructs can be built.  For 
example, the quarks $q$ are coloured and so can form hadrons by colour
confinement.  Then these hadrons have (soft) nuclear forces between them, and 
can combine via such forces to give complex nuclei.  Further, some of these 
nuclei being charged, they can combine by electromagnetic interaction with 
charged leptons to form atoms and molecules, and eventually us.  Similar 
constructs are in principle possible also with the co-quarks $Q$ and co-leptons 
$L$, forming a co-sector which is possibly as complex and as vibrant in 
interactions within itself as our own standard sector.  

\end{itemize}

This dichotomy is illustrated in Table \ref{DoM}.  Notice that the hidden 
sector there is labelled ``hidden'' only because we have not seen in experiment 
any of the particles listed in that sector.  But we have yet to understand 
theoretically in FSM why it is that we have not seen them, namely again the
answer to {\bf [Q1]}.  This is a task that we shall have to come back to later.

\begin{table}
  \begin{tabular}{l|l|l}
    & {\bf Standard Sector} & {\bf ``Hidden Sector''} \\
    \hline 
    Building blocks & $(h), (W,Z), (q,\ell)$ & $H, G, F$ \\
    & \quad point-like, perturbative & \quad point-like, perturbative
    \\
    & \quad interactions & \quad interactions \\
    \hline
    Bound states of above & $q \bar{q}$: mesons (bosons) & \\
    \quad by colour confinement & $qqq$ baryons (fermions) & \\
    & \quad bulky, non-perturbative &\\
    & \quad soft interactions &\\
    \hline
    Bound states of above & & $Q \bar{Q}$ co-mesons (bosons) \\
    \quad by flavour confinement & & $QQ$ co-baryons (bosons)\\
    && \quad bulky, non-perturbative\\
    && \quad soft interactions??\\
    \hline
    Bound states &nuclei & co-nuclei? \\
    \quad by soft interactions && \\
    \hline
    Bound states by e.m. & atoms, molecules, \ldots, us & co-atoms, co-molecules?
    \\
    \hline
  \end{tabular}
  \caption{The Dichotomy of Matter (according to the FSM)}
  \label{DoM}
  \end{table}


\section{Transfer of technology between sectors}

Now, it is unexpected, indeed surprising, that the FSM, which was constructed 
originally to understand the generation puzzle for quarks and leptons in our
sector should have led us to such a dichotomy with a ``hidden sector''.  But 
once there, the ``hidded sector'' would be far from unwelcome when we recall 
that more than half our world is made up of dark matter the nature of which is 
still hidden from us \cite{pdg1}.  
However, merely to suggest the existence of such a
sector is not much use unless one can also suggest the means for studying it, 
to explain why it should be hidden in the first place, and then, having done 
so, to find ways to probe into its hidden secrets.  And most gratifyingly, the 
parallel depicted in Figure \ref{fvsc} also allows us to do so via what we 
might call:
\begin{itemize}   
\item {\bf [ToT]} {\it Transfer of technology between the standard and 
``hidden'' sectors}, meaning that the same machinery used to investigate 
$h$, $W, Z$ and $q, \ell$ on the left of Figure \ref{fvsc} can be applied 
also to their counterparts $H$, $G$, and $F$ on the right, allowing one to 
explore the ``hidden sector'' in depth.\footnote{Although the ``hidden sector''
being still unknown, the transfer of technology will mostly go one way at 
present, namely from the known standard sector to the other, the transfer 
can go in principle the other way also.  Indeed, in \cite{tfsm} one had 
already an example where it was from the study of radiative corrections 
to the $F$ self-energy in the ``hidden sector'' that the RGE for the rotation 
of $\balpha$ was derived, which was what gave the result for quarks and 
leptons in Table \ref{tfsmfit}.}   
\end{itemize}

In particular: 
\begin{itemize}
\item  The perturbative method used so succesfully to study the properties and 
interactions of the weak particles $h$, $W, Z$, and $q, \ell$ in the standard
sector should be applicable also to their analogues $H$, $G$, and $F$ in the
hidden sector.
\end{itemize}  
For the weak particles in the flavour theory, perturbation theory is usually 
carried out in the symmetry-breaking picture, but the calculation is basically 
the same in the confinement picture we favour, only interpreted differently, as 
has been shown by 't Hooft \cite{tHooft}, and Banks and Rabinovici \cite{Bankovici}.
Further, according to 't Hooft, the perturbative method in the electroweak 
theory is permissible if the vacuum expectation value
of the scalar Higgs field $\zeta_W (\sim 246$ GeV) is large.  
So it should be permissible also in the colour theory in FSM where the
vacuum expectation value of 
the scalar framon is even larger $\zeta_S \sim$ TeV.

If this is true, it leads immediately to the following results:
\begin{itemize}
\item It post-justifies the one-loop calculation carried out in \cite{tfsm}
which gave the result in Table \ref{tfsmfit}.  This calculation was initially 
carried out merely as the simplest one could do, without it being then 
considered whether a perturbative approach was in fact justified.
\item It justifies the entry in Table \ref{DoM} that the particles $H, G, F$
are point-like.  This was our conclusion for the weak particles $h$, $W, Z$, 
$q, \ell$ in the flavour theory because perturbation theory applied, and so
by the same token, the conclusion should hold also for the particles $H, G$ 
and $F$ in the colour theory.  Now, previously, in \cite{cfsm} an intuitive 
physical argument was suggested to support what was then a conjecture that the
$H, G, F$ are point-like for having little soft interactions based on the fact 
that their framon constituents have short life-times.  This is now seen to be 
unnecessary since the point-likeness of the $H, G, F$ is supposedly already
guaranteed by their perturbative nature.  By dispensing then with that argument 
as an assumption, one has put the conclusion on a firmer basis and confirms 
the parallel drawn up in Figure \ref{fvsc} above.  However, this does not by 
itself invalidate that qualitative argument, which may still be retained as a 
useful intuitive picture of why quark bound states via colour confinement 
(hadrons) have soft interactions while framonic bound states via the same 
colour confinement have none.
\item It opens up a huge vista of coming explorations of the hidden sector 
using perturbative methods which can in principle rival in detail and
complexity that for our standard sector and turn into a major industry.
\end{itemize}  

\section{$G$-modified mixing}

The last item listed is of course exciting, but before we get carried away,
let us first perform the initial steps to test whether such a programme is at all 
sensible and likely to bear fruit.  As a first step, let us investigate the 
mass spectra and interaction vertices of the $H$s, $G$s, and $F$s.  As in the 
familar flavour theory, these can be obtained by expanding the framon action in 
fluctuations of the framon about its vacuum expectation value.
This one can do when one recalls 
that the framon action is strongly constrained by the requirement of the double 
invariance under $G \times \tilde{G}$ essentially fixing
its form, only 
dependent on some parameters.  Though lengthy and cumbersome, the calculation 
is fairly straightforward and is reported in \cite{cfsm}.  As an example, only 
that for the mass spectrum of the $G$s is outlined here, which is in some sense 
the simplest but has particular physical significance.  

The mass matrix for the vector bosons is obtained by expanding the kinetic 
energy term of scalar framon fields in their fluctuations about their vacuum 
values to leading order, meaning in this case the substitution of the framon 
fields by their vacuum values.  We are familiar in the standard electroweak 
theory how the calculation there gives diagonal masses to the $SU(2)$ fields 
$B^i_\mu$ labelled by the Pauli matrices $\tau_i$ to give the massive vector 
bosons $W^i$, except for the third component $B^3_\mu$ which mixes with the 
$U(1)$ field $A_\mu$ to give the electromagnetic field $\gamma_\mu$ and $Z_\mu$. 
Not surprisingly, the same machinery applied to the FSM, extended to include 
both the flavour and colour sectors, gives a mass matrix for the vector bosons 
which is still diagonal when the $SU(2)$ fields remain labelled by the Pauli 
matrices while the $SU(3)$ fields are labelled by the Gell-Mann matrices 
$\lambda_k$, except now for an extended and modified mixing among $A_\mu, 
B_\mu^3$ and $C_\mu^8$. 

Because of this latter mixing, two critical questions immediately arise.

First,
\begin{itemize}  
\item  {\it Will the photon remain massless?}  Or else the FSM theory will be 
ruled out right away given that the photon mass has already been checked to 
hardly disputable accuracy.
\end{itemize}  
Fortunately the answer is yes: one can keep the photon massless if one chooses 
the electric charges of the colour framons judiciously, as Salam and Weinberg 
did for the flavour framon (Higgs scalar) when constructing the electroweak 
theory.  The appropriate choice here in the colour case is $-\third, -\third, 
+\twothirds$ respectively for the 3 columns of $\bPhi$ in {\bf [CF]}, where
the third column is that aligned with $\balpha$ and the first and second 
columns are perpendicular to it \cite{cfsm,zmixed}.  This choice is similar 
to the Salam-Weinberg theory of the charges $-\half, +\half$ for the two 
columns of $\Phi$ in {\bf [FF]} except that in this case, the orientation of
the columns need not be specficied, basically again because of (\ref{minemb}).  
These charges were in fact needed to complete the specification of the framon
representations in $G \times \tilde{G}$ but were missed out deliberately before
in {\bf [FF]} and {\bf [CF]} because they could not then be specified.  Now 
that they are settled, they can be inserted there for completeness.  

Secondly,
\begin{itemize}
\item {\it What about the new mixing for the $Z$?}  The extra mixing above
will mean deviations from the standard Weinberg mixing scheme which has 
already been tested to great accuracy in experiment.  Will these deviations 
then lead to violations of present experimental bounds? 
\end{itemize} 
This question cannot be answered yet in full given that the FSM has not been 
developed sufficiently for loop diagrams (radiative corrections) to be 
calculated in general, which will be needed for checking with experiment in 
depth.  However, initial tests can be devised as follows.  Assuming, as is 
generally accepted, that the SM agrees with experiment to within present 
bounds, and that the deviations of the FSM from the SM in loop corrections 
are of higher order, we compare the tree-level results of the FSM and the SM 
and if the difference is within experimental bounds, we conclude that the FSM 
results are also within present experimental bounds.  This criterion has been
applied in \cite{zmixed} to the 
following most urgent cases:
\begin{itemize}
\item {\bf (a)} $m_Z - m_W$,
\item {\bf (b)} $\Gamma(Z \rightarrow \ell^+ \ell^-)$,
\item {\bf (c)} $\Gamma(Z \rightarrow q \bar{q})$
\end{itemize}
where the deviations at tree level of the FSM from the SM are evaluated.  These
all depend on $m_G$, the mass of a new vector boson $G$ (or equivalently the
vacuum expectation value
of the colour framon $\zeta_S \sim 2 \times m_G$) as the only parameter. 
And it is found that so long as $m_G > 1$ TeV, then the FSM deviations from the 
SM at tree level will all remain inside the present already very stringent 
experimental bounds \cite{zmixed}.  There are some subtle cancellations which 
have allowed this to happen and point perhaps to a deeper reason for the 
agreement not yet understood.

\begin{figure}
\centering
\includegraphics[scale=0.2]{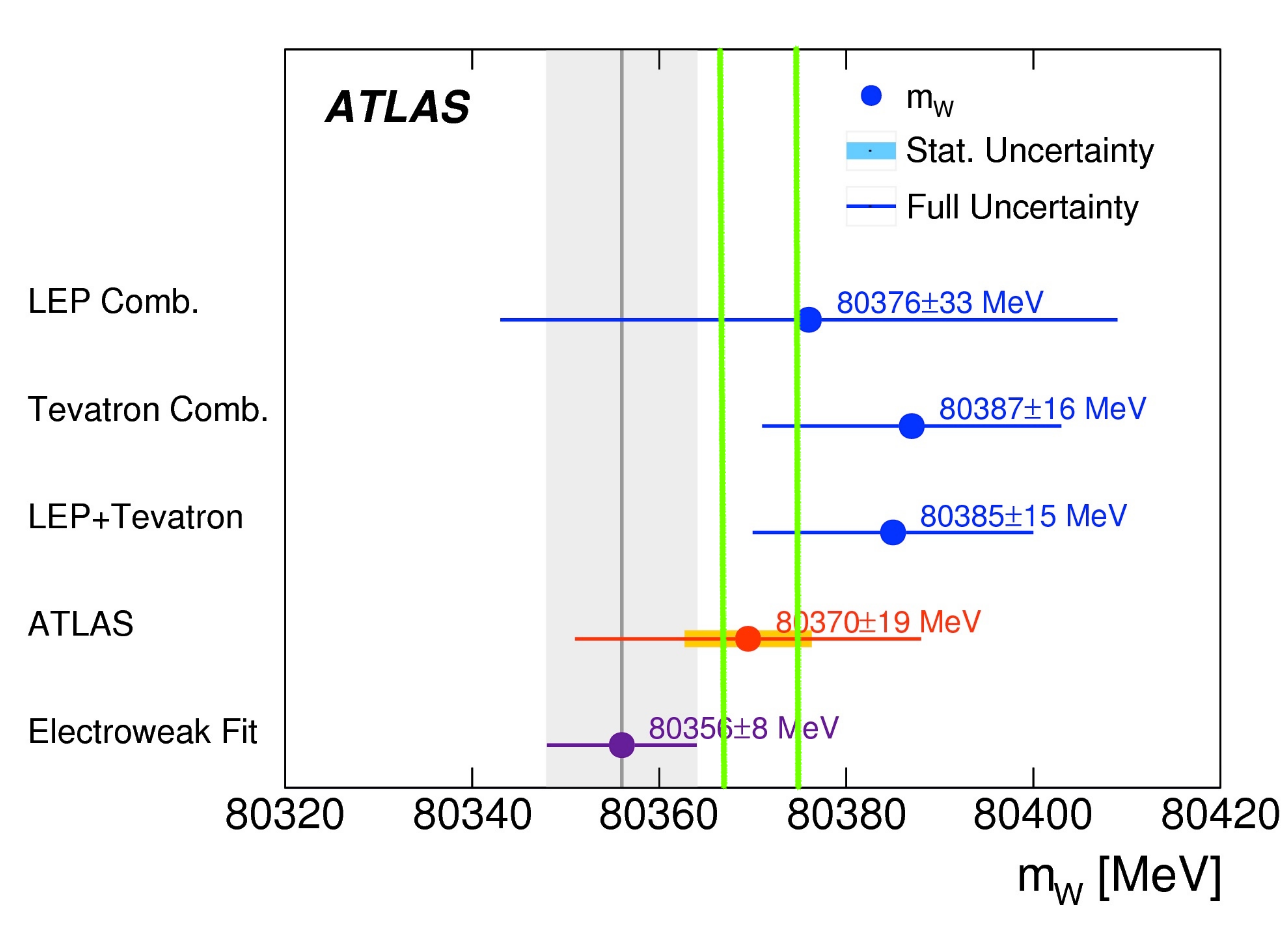}
\caption{The ATLAS measurement of the W boson mass and the combined values measured at the LEP and Tevatron colliders compared to the Standard Model prediction (mauve) and the FSM predictions (green) at $\zeta_S = 2.0$ TeV (left) and $\zeta_S = 1.5$ TeV (right).}
\label{Wmass}
\end{figure}

One can turn the argument around, of course, and treat these deviations as new
physics to be tested when experimental accuracy further improves.  In this
direction, an observation on the $W$ mass may be noteworthy.  The FSM predicts,
via the extra mixing with the new boson $G$, a smaller value for the mass shift
$m_Z - m_W$ than the SM.  This means that starting with the better measured
$m_Z$ to predict $m_W$, as it is usually done, one would obtain a larger value
for $m_W$ in the FSM than in the SM.  The present experimental situation as
recently summarized by ATLAS \cite{mWATLAS} is shown in Figure
\ref{Wmass}  where
it is seen that successive measurements at LEP, the Tevatron, and the LHC all
actually give central values for $m_W$ bigger than the SM prediction, as 
the FSM suggests, although the excess is only 1--2 $\sigma$, and therefore 
statistically not yet significant.  But if in future experimental accuracy for 
$m_W$ further improves, then it would be meaningful to ask whether the excess
predicted by the FSM really exists.  The FSM prediction depends on $\zeta_S$,
the vacuum expectation value of the colour framon which works out to be about 
$2m_G$.  In Figure \ref{Wmass}, the FSM prediction is shown in green for 
$\zeta_S = 2, {\rm and}\ 1.5$ TeV, (corresponding to  $m_G \sim 1.0\ {\rm and}\ 
0.75$ TeV respectively).  The FSM predictions actually seem to give a better 
fit than the SM to the present available data.

The deviations found in \cite{zmixed} of the FSM from SM in the decay widths
of $Z \rightarrow \ell^+ \ell^-$ and $Z \rightarrow q \bar{q}$ may also appear 
as new physics when experimental accuracy improves.  Obviously, however, as 
far as the change in mixing of the vector bosons ($G$-modified mixing) from 
the SM to the FSM is concerned, the prime new physics target would be the 
discovery of the vector boson $G$ itself.  We shall postpone discussions of 
this till later when further facts are known.

\section{Masses and interactions of the $H$, $G$ and $F$}

Let us turn back now to the exploration of the ``hidden'' sector, continuing 
with the mass spectrum of the $G$s.  We noted already that their mass-squared 
matrix is nearly diagonal except for the mixing of $G_8$ already studied.  The 
diagonal values are as listed below:
\begin{eqnarray}
& \frac{1}{6} g_3^2 \zeta_S^2 (1 - R) & {\rm for}\ K = 1, 2, 3; \nonumber \\
& \frac{1}{12} g_3^2 \zeta_S^2 (2 + R) & {\rm for}\ K = 4, 5, 6, 7; \nonumber \\
& \frac{1}{6}  g_3^2 \zeta_S^2 (1 + R) & {\rm for}\ K = 8,
\label{MGsq}
\end{eqnarray}
where $g_3$ is the colour coupling and $\zeta_S$ the vacuum expectation value
of the colour framon.  They are very similar to those in the familiar flavour 
case except for the appearance of factors depending on a parameter $R$.  This 
parameter is a ratio which measures the relative strengths of symmetry-breaking 
versus symmetry-restoring terms in $\widetilde{SU}(3)$ which has no analogue 
in $\widetilde{SU}(2)$ because of the condition (\ref{minemb}) imposed on the
flavour framon for reasons already stated.

\begin{figure}
\centering
\includegraphics[scale=0.42]{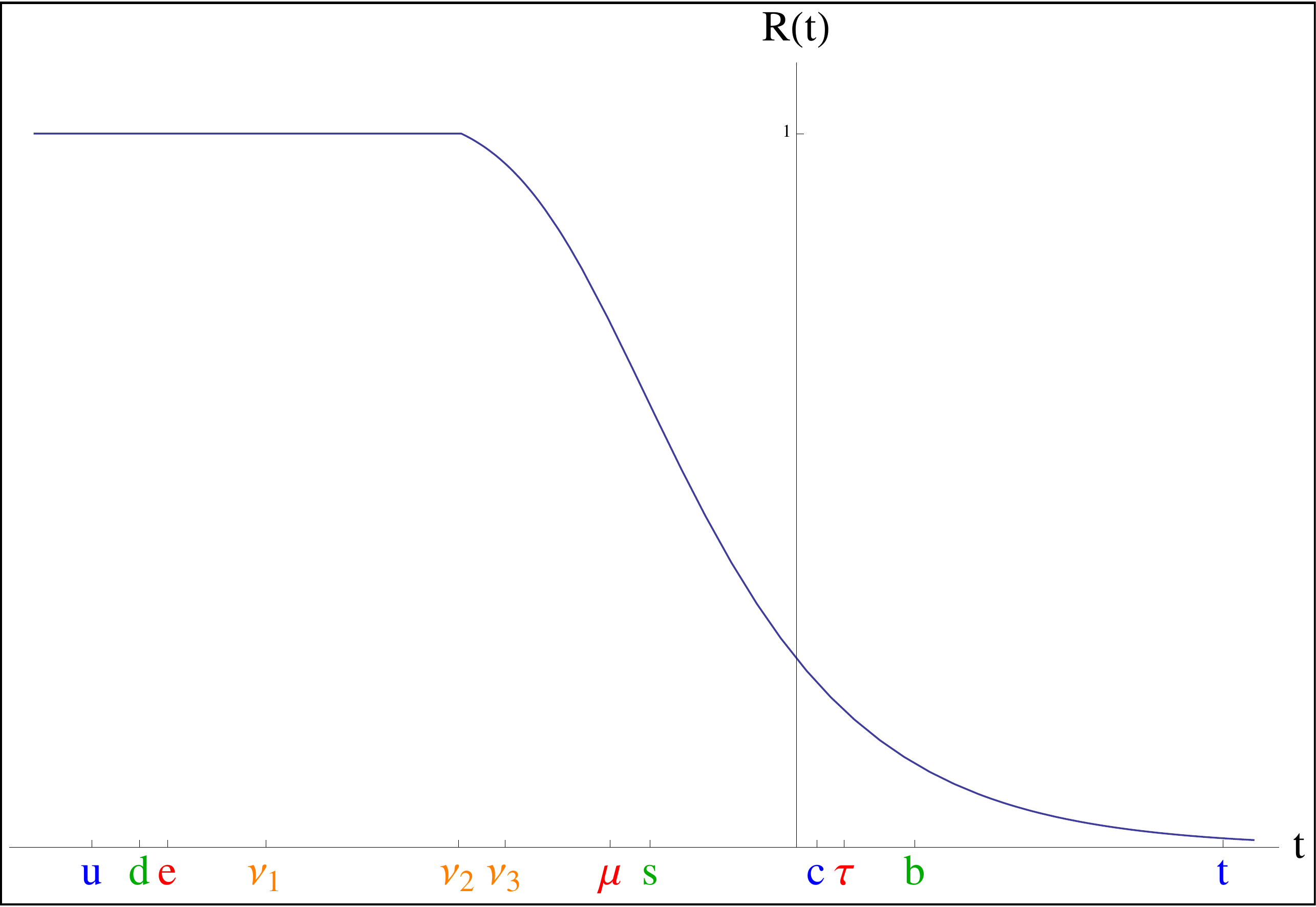}
\caption{Dependence of $R$ on scale $\mu$ \cite{tfsm}}
\label{Rtfsm}
\end{figure}

Now this ratio $R$ figures prominently in the RGE for the rotation of $\balpha$
which was used in \cite{tfsm} to give the fit in Table \ref{tfsmfit}.  Since
both these quantities are connected with how the symmetry $\widetilde{SU}(3)$
is broken, it is not surprising that their scale-dependences are correlated.
Thus, corresponding to the trajectory of $\balpha$ of Figure \ref{florosphere},
one has obtained in \cite{tfsm} the scale-dependence of $R$ shown in Figure
\ref{Rtfsm}.  Given that this dependence is quite as strong as that of $\balpha$
which was used to fit the quark and lepton spectra, it seems reasonable to take
account also of the parallel scale-dependence of $R$ in studying the spectra of
the $G$s.  In that case, the matrix elements in (\ref{MGsq}) depend on scale 
and we have to question at what scale or scales are the physical masses of the
$G$s are to be evaluated.     

The general consensus is that when the mass matrix is scale-dependent in QFT
the physical masses of particles are to be evaluated each at their own mass 
scales, meaning that they are each to be solutions of the equation:
\begin{equation}
m_x(\mu) = \mu,
\label{physmass}
\end{equation}
where $m_x(\mu)$ is the scale-dependent eigenvalue of the mass 
matrix corresponding to the particle $x$ under consideration. 

Apply now this criterion to the $G$s.  From
Figure \ref{Rtfsm}, we see that  $R$ is 
small, $0.02$ or less, for $\mu > m_Z$, so that the eigenvalues in (\ref{MGsq}) 
are all nearly degenerate.  Further, from the analysis given above for the 
change in mass shift $m_Z - m_W$ due to $G$-modified mixing, $\zeta_S > 2$ TeV, 
which means that the physical masses obtained from (\ref{physmass}) will be in 
excess of 1 TeV and nearly degenerate, apart from two exceptions.  First, $G_8$ 
will be pushed higher by a small amount via mixing with $\gamma$ and $Z$ to 
form $G$.  Secondly, and much more dramatically, the mass matrix elements of 
$G_1, G_2, G_3$ in (\ref{MGsq}) all carry a factor $1 - R$ which according to 
Figure \ref{Rtfsm} vanishes at $\mu$ around 17 MeV.  This means that the equation
(\ref{physmass}) is bound to have another solution just above 17 MeV.  Now it 
has been suggested in \cite{tfsm} in a parallel case for leptons and quarks 
that whenever a second lower solution exists for (\ref{physmass}), then it 
should be taken as the physical solution since the higher solution will be 
unstable against decay into the lower.  Indeed, this assertion is what gives 
the result $m_u <m_d$ much vaunted in {\bf [b]} above.  If it is accepted here 
also, then we have the spectrum for $G$s summarized in Table \ref{Gmass}.
Notice the characteristic separation of the spectrum into two groups, one, say
$G_{\rm heavy}$ with masses of the order TeV, and the other $G_{\rm light}$ with masses
$\sim 17$ MeV, where those in $G_{\rm light}$ are all electrically neutral, which
will be of significance for discussions later.\footnote{However, a serious word 
of caution is needed for the 17 MeV prediction for the mass of $G_{\rm
  light}$, and 
later on for $H_{\rm light}$.  We recall that its derivation is based on strict 
adherence to the following 3 criteria: (i) physical mass of a particle is to 
be measured at its own mass scale, i.e. solution of (\ref{physmass}), (ii) 
validity of the fit in \cite{tfsm} giving Table \ref{tfsmfit}, and (iii) when 
there are two solutions to (\ref{physmass}), one takes the lower.  None of 
these three theoretical criteria is beyond reasonable doubt.  Thus with these 
alone to deduce a mass of 17 MeV from a system with a natural scale of order 
TeV is a little audacious.  On the other hand, for phenomenological support,  
one can quote, first, the apparent success in Table \ref{tfsmfit}, where using 
the same three criteria, one gets right the lightest generation properties: 
(i) $m_u, m_d, m_e$ all of order MeV, (ii) $m_u < m_d$, both of which facts will 
otherwise be very hard to understand.  Secondly, there is also the coincidence
in mass with the Atomki anomaly to be mentioned later.  How much trust one can
put on this prediction is a question which needs to be asked again after more
consideration.}.

\begin{table}
\center
\begin{tabular}{|l|l|l|}
\hline
Particle & State & Mass \\ 
\hline \hline
$G^0$ & mixture of $G_8, Z$ and $\gamma$ & $\geq 1.1\ {\rm TeV}$ \\
\hline
$G^+$ & $\frac{1}{\sqrt{2}}[G_4 + i G_5]$ & {} \\
$G^-$ & $\frac{1}{\sqrt{2}}[G_4 - i G_5]$ & {} \\
{}    & {} &      $\geq 1.0\ {\rm TeV}$ \\
$G^{'+}$ & $\frac{1}{\sqrt{2}}[G_6 + i G_7]$ & {} \\
$G^{'-}$ & $\frac{1}{\sqrt{2}}[G_6 - i G_7]$ & {} \\
\hline
$G_1^0$ & {} & {} \\ 
$G_2^0$ & {} & $ \sim 17\ {\rm MeV}$ \\
$G_3^0$ & {} & {} \\
\hline \end{tabular} 
\caption{Suggested spectrum of the $G$ states} 
\label{Gmass}
\end{table}

The mass spectra of the $H$s and $F$s have been similarly investigated, for
the $H$s by expanding the framon potential and for the $F$s the Yukawa terms.
Apart from some parameters with precise values yet unknown, the $H$ spectrum
poses little problem, dividing as did that for the $G$s into two groups, one
$H_{\rm heavy}$ with masses probably in excess of TeV, and one $H_{\rm
  light}$ with mass
$\sim 17$ MeV.  Again the members of $H_{\rm light}$ are all electrically neutral.  
The spectrum for the $F$s, however, is more problematic because, with no known
geometrical significance for the fermion fields, one is unsure what the
fermion fields are
which should enter into the Yukawa couplings.  In addition,
the co-neutrinos among the $F$s can be affected by a see-saw mechanism \cite{seesaw} similar
to that for neutrinos in the standard sector, and so may end up with masses
considerably lower than may appear at the tree level.  The spectrum for $F$s
suggested in \cite{cfsm} is thus model-dependent, standing only to be adjusted 
when more information becaomes available.  As it stands, however, it divides
also into a $F_{\rm heavy}$ and a $F_{\rm light}$ group as do the
spectra for $G$ and $H$.

Expanding further the framon action to higher orders, one obtains interaction
vertices for the $H$s, $G$s and $F$s.  This is done in \cite{cfsm} giving some 
10 pages of vertices between the various $H$s, $G$s, and $F$s.  An outstanding 
feature in these results is that despite the many vertices coupling the $H$s, 
$G$s and $F$s among themselves, there are very few which link them to the weak 
particles $h$, $W, Z$, and $q, \ell$ in the standard sector.  Indeed, the only 
examples found for the latter type of couplings are due to the mixing already
mentioned of $G_8$ with the $\gamma$ and $Z$ of the standard electroweak theory 
and a similar but somewhat more complex mixing of certain $H$ states with the 
electroweak theory Higgs state $h_W$.  This lack of couplings linking the two 
sectors comes about mainly because of the choice (\ref{framonrepl}) for the 
representation of the framon made at the beginning.  For instance, this 
choice implies that the framon kinetic energy term, which is the most prolific 
in spawning vertices, is a sum of two terms, one for the flavour and one for the 
colour sector, with only the $U(1)$ gauge field $A_\mu$ linking the two.  This 
lack of couplings betwen the two sectors will figure prominently later when 
physical consequences are considered.

\section{Exploring the hidden sector---a few first steps}

Given the mass spectra of the $H$s, $G$s and $F$s and the interaction vertices 
between them, one has then the basic ingredients to develop the perturbaton 
theory for the $H$s, $G$s and $F$s in the hidden sector.  There may, of course,
be unforeseen difficulties, but the perturbation theory can in principle 
be developed to a similar degree of sophistication as for the standard sector 
including loops for radiative corrections and so on.
In this perspective, therefore,
what has been done so far has merely scratched the surface of what is accessible,
and not so evenly even at that.  At tree level, little is examined beyond the 
mass spectra and interaction vertices of the $H$s, $G$s and $F$s.  And as for 
loops, the only venture made so far has been that reported in \cite{tfsm} 
giving the RGE for the rotation of $\balpha$.  Nevertheless, it is worthwhile
to pause awhile for breath and take stock of the physical situation revealed by
that little which has been found. 

Our first task is to return to the question {\bf [Q1]} why we have all along 
been unaware of the particles $H$, $G$ and $F$, or why the hidden sector has so 
far been hidden from us.  Let us see whether we can now venture an answer.  The 
analysis of the mass spectra and interaction vertices above has shown us that 
the hidden sector is as heavily populated and as vibrant in interactions within 
itself as our standard sector, in fact perhaps even more so.  However, there 
are only restricted communications between the two sectors so that we who live 
in the standard sector may have difficulty knowing about what we call the 
``hidden sector", as they also who abide in the other sector may have difficulty 
knowing about us.  Nevertheless, some of the $H$s, $G$s and $F$s, such as $G^0$ 
and $G^\pm, G^{'\pm}$, would still manifest themselves, if they exist; thus $G^0$ 
by decaying into $\ell^+ \ell^-$, for example, and the other charged states by 
exchanging a photon with our standard charged particles.  However, one might 
argue that these particles being heavy with masses of order TeV would have 
decayed away by our epoch even if they were present in the early universe so 
that none will occur now naturally.\footnote{This is assumed to be the case
even though, with insufficient knowledge of the $F$ spectrum, it has not been
possible to work out the decay process of the $H$s, $G$s and $F$s in every 
case.}  If we wish to see them, we shall have to produce them in the laboratory,
 and this is not easy, given their very high masses and their reluctance in 
coupling to normal matter with which we do our experiments.  What can occur 
naturally are the low mass states, that is,  what we call the $H_{\rm light}$, 
$G_{\rm light}$ and $F_{\rm light}$ some of which are likely to be stable.  But these 
being electrically neutral, with little interaction with ordinary matter, will 
behave like dark matter to us.  One can then claim for these reasons that the 
whole sector has so far been hidden from us, answering thereby {\bf [Q1]}.

Suppose this is true, then knowing now how the particles in the hidden sector
hide themselves, can we not find some gaps in their defence to get at them?
Based just on what is known, one can suggest the following:
\begin{itemize}
\item  The most promising is probably the vector boson $G$.
If the tempting indications of Figure \ref{Wmass} are taken seriously, the $G$ 
mass would not be much above 1 TeV.  Its known mixing with $\gamma$ and $Z$ 
means that it will decay into $\ell^+ \ell^-$ pairs with a known width 
(depending only on its mass) which is already calculated \cite{zmixed}. 
Its production cross section at LHC seems calculable and is under investigation.
Its total width depends on the projected mass spectra and couplings of the 
$H$, $G$ and $F$ which are already available though not yet fully tested. 
With these bits of information, one may soon be able to make practicable
suggestions for $G$ to be searched for as a $\ell^+ \ell^-$ bump at the LHC.  
Its discovery would not only serve as a detailed check of the FSM scheme but 
also, according to this, a window into the ``hidden'' sector, for it will be 
into the dark matter candidates ($G_{\rm light}$ and $H_{\rm light}$ states of mass 
around 17 MeV or co-neutrinos of mass $< 8$ MeV) that $G$ will mostly decay.
\item  The next most promising is probably $G_3$, with predicted mass 17 MeV,
which can decay into $e^+ e^-$ via  a photon attached to a framon loop.  Now,
it so happens that coincidentally at this predicted mass, an anomaly has been 
reported by the Atomki collaboration \cite{atomki} in the decays: $Be_8^* 
\rightarrow Be_8 + e^+ e^-$.  Not enough  work has  been done yet on $G_3$ to
ascertain whether it can explain the Atomki anomaly, and the anomaly itself
needs to be independently confirmed, but the coincidence is intriguing.  One
notes that the prediction of 17 MeV is deduced from the result of \cite{tfsm}
which predates the Atomki report of the anomaly. (Note, however, the remark in
footnote 5.)  In any case, independently of Atomki, it is worthwhile studying 
further this mass region around which there has already been a lot of activity, 
prompted by such phenomena as the $g - 2$ anomaly \cite{gminus2} or the proton radius puzzle \cite{protonradius}.
\item The remaining states $G_1, G_2$ in $G_{\rm light}$ (and their counterparts in
$H_{\rm light}$) at 17 MeV appear stable, unless some co-neutrions can acquire masses
via a seesaw mechanism low enough for these to decay into.  In any case, there
will be dark matter candidates at these low masses for experiment to search for.
They are probably beyond the reach of LUX \cite{lux}, LZ \cite{lz},
and other experiments 
which concentrate on the multi-GeV region but may be accessible with new
techniques such as SENSEI \cite{sensei}.  Although light, these dark candidates
are expected to occur in abundance. First, being binary objects, they would be statistically
easier to form in the early universe than baryons which would require
the coincidence of three quarks for their formation.  Secondly, these
dark matter candidates occur frequently as decay products of higher
$H, G, F$ states.  And so, though light, they may make up a sizeable
fraction of the missing mass.
However, much 
more work will be needed to ascertain whether they may make up the bulk of it.
One novel feature of these particles as dark matter is that they are predicted
by a model not specifically designed
for dark matter itself but one constructed for another purpose,
and which model already prescribes a great deal about how these particles will 
behave, given that the action is known and perturbative calculations apply.
\item  The charged particles in the ``hidden'' sector couple to the photon as
usual so that these could be pair-produced in a $e^+ e^-$ collider provided the 
energy is high enough.  They will give rise to step increases of the ration
$R$ as usual in $e^+e^-$ collisions.  Now the model Yukawa coupling proposed 
tentatively in \cite{cfsm} suggests masses of these charged $F$s of the order
TeV, in which case they would be above the $e^+ e^-$ colliders being planned.
But this may not be the case for other choices of the fermion fields in 
constructing the Yukawa couplings.
\end{itemize}

In brief, it is unexpected, indeed rather amazing, that constructed
originally for addressing the generational puzzle in our own standard
sector, the FSM has provided us, in addition, with a lead into the
mysterious dark matter world.  It is made even more intriguing by the
fact that, the physics of the two sectors being closely interlocked,
what is known about our standard sector provides also the technology
for exploring the other.  And the first few steps along the indicated
direction have already yielded some items of new physics testable by
experiment.  Thus, if there is truth in what is said, we may soon be
probing into a vast new world which has so far been hidden from us.

\vspace{.5cm}

We are indebted to Paul Hoyer for suggesting this descriptive summary.

\end{document}